\documentstyle[12pt]{article}

\def\title #1 {
   \headsep=1.0in
   \baselineskip=30pt
                        \begin{center}
   {\titlebold #1}
   \end{center}
                        \vskip .75in }
\def\author #1 {
   \baselineskip=30pt
   \begin{center}
   {\timeslarge #1}
   \end{center}
                        \vskip .25in }
\def\address #1 {
   \baselineskip=24pt
   \begin{center}
   {\timesitalic #1}
   \end{center}
   \vskip 1.0in }

\def\d {\mbox{$\delta$}}

\def\half {\mbox{$\textstyle {1 \over 2}$}}

\def\disp {\displaystyle}

\def\conj #1 {\overline #1}

\def\be {\begin{equation}}
\def\ee {\end{equation}}

\def\ba {\begin{array}}
\def\ea {\end{array}}
\def\bea {\begin{eqnarray}}
\def\eea {\end{eqnarray}}

\def\et {$$}

\def\etn {$$}

\def\ett {$$}

\def\ettn{$$}
\def\non  {\nonumber}
\def\eqn#1 {(\ref{#1}) }

\newdimen\twoeqncolwidth
\newdimen\twoeqncolwidtha
\newdimen\twoeqncolwidthb
\newdimen\twoeqncolsep
\newdimen\twoeqnlinset
\def\twoeqn#1&#2\et{
   \hbox to\twoeqnlinset{\hfil}
   \hbox to\twoeqncolwidth{$\disp#1$\hfil}
   \hbox to\twoeqncolsep{\hfil}
   \hbox to\twoeqncolwidth{$\disp#2$\hfil}\eqno{\rm (\theequation)}$$}
\def\twoeqnt#1&#2\ett{
   \hbox to\twoeqnlinset{\hfil}
   \hbox to\twoeqncolwidtha{$\disp#1$\hfil}
   \hbox to\twoeqncolsep{\hfil}
   \hbox to\twoeqncolwidthb{$\disp#2$\hfil}\eqno{\rm (\theequation)}$$}
\def\twoeqnn#1&#2\etn{
   \hbox to\twoeqnlinset{\hfil}
   \hbox to\twoeqncolwidth{$\disp#1$\hfil}
   \hbox to\twoeqncolsep{\hfil}
   \hbox to\twoeqncolwidth{$\disp#2$\hfil}\eqno\phantom{\rm (\theequation)}$$}
\def\twoeqntn#1&#2\ettn{
   \hbox to\twoeqnlinset{\hfil}
   \hbox to\twoeqncolwidtha{$\disp#1$\hfil}
   \hbox to\twoeqncolsep{\hfil}
   \hbox to\twoeqncolwidthb{$\disp#2$\hfil}\eqno\phantom{\rm (\theequation)}$$}

\def\rawpicture #1 by #2 (#3){
  \vbox to #2{
    \hrule width #1 height 0pt depth 0pt
    \vfill
    \special{picture #3} 
    }
  }
\def\scaledpicture #1 by #2 (#3 scaled #4){{
  \dimen0=#1 \dimen1=#2
  \divide\dimen0 by 1000 \multiply\dimen0 by #4
  \divide\dimen1 by 1000 \multiply\dimen1 by #4
  \rawpicture \dimen0 by \dimen1 (#3 scaled #4)}
  }
\renewcommand{\theequation}{\thesection.\arabic{equation}}
\renewcommand{\title}[1]{\large\bf \mbox{}\\ \mbox{}\\ \mbox{}\\
     #1\bigskip\medskip\\}
\renewcommand{\author}[1]{\large #1\\ \smallskip}
\renewcommand{\address}[1]{{\narrower\normalsize\it #1\\}\bigskip}
\def\wt#1#2#3#4#5#6{#1\!\!\mbox{
$\left(\left.\matrix{#5&#4\cr#2&#3\cr}\right|\mbox{$#6$}\right)$}}
\def\ade{$A$--$D$--$E$\space}
\long\def\ignore#1{}
\def\olambda{\lambda}

\def\tiny{\small}
\def\be{\begin{eqnarray}}
\def\ee{\end{eqnarray}}

\def\and{\;\;{\rm and }\;\;}
\def\disp{\displaystyle}

\def\1{\mbox{\boldmath $I$}}

\def\){\right)} \def\({\left(\stackrel{}}
\thicklines

\def\sqface#1#2#3#4#5#6#7{\rule[-2.8\unitlength]{0in}{5.6\unitlength}
\begin{picture}(4,4)(-#6,-#7)
\put(0,-2){\vector(1,0){4}}
\put(4,2){\vector(0,-1){4}}
\put(0,2){\vector(1,0){4}}
\put(0,2){\vector(0,-1){4}}
\put(0,-3.2){\makebox(0,0)[b]{\small \mbox{$#1$}}}
\put(4,-3.2){\makebox(0,0)[b]{\small \mbox{$#2$}}}
\put(4,2.5){\makebox(0,0)[b]{\small \mbox{$#3$}}}
\put(0,2.5){\makebox(0,0)[b]{\small \mbox{$#4$}}}
\put(2,0){\makebox(0,0){\small \mbox{$#5$}}}
\end{picture}}

\newcommand{\nc}{\newcommand}

\nc{\sm}[1]{{\scriptstyle #1}}
\nc{\ssm}[1]{{\scriptscriptstyle #1}}
\nc{\vect}[1]{\mbox{\boldmath$#1$}}

\nc{\pos}[2]{\makebox(0,0)[#1]{$#2$}}
\nc{\spos}[2]{\makebox(0,0)[#1]{$\sm{#2}$}}
\nc{\text}[6]{\begin{picture}(#1,#2)\put(#3,#4){\pos{#5}{\displaystyle#6}}
\end{picture}}

\nc{\doublerightfacerow}[4]{\begin{picture}(9.4,2.4)(-0.2,-1.2)
\put(0,-1){\ltri}
\put(1,-1){\rect}\put(2,-1){\laddash}\put(3,-1){\rect}
\put(4,-1){\diam}\put(6,-1){\diam}\put(8,-1){\rtri}
\put(8.62,0){\spos{}{#1}}\put(7.05,0){\spos{}{#2}}\put(0.4,0){\spos{}{#4}}
\put(5,0){\spos{}{#3}}
\put(1.5,-0.5){\spos{}{#1}}\put(3.5,-0.5){\spos{}{#1}}
\put(1.5,0.5){\spos{}{#4}}\put(3.5,0.5){\spos{}{#4}}
\end{picture}}

\nc{\gridfour}{\begin{picture}(6,5)(1,1)
\multiput(0,0)(0,1){4}{\line(1,0){4}}
\multiput(0,0)(2,0){3}{\line(0,1){3}}
\put(1.05,0.5){\spos{}{0}}\put(3,0.5){\spos{}{1/2}}
\put(1,1.5){\spos{}{1/16}}\put(3,1.5){\spos{}{1/16}}
\put(1,2.5){\spos{}{1/2}}\put(3.05,2.5){\spos{}{0}}
\put(1.05,-0.5){\spos{}{1}}\put(3.05,-0.5){\spos{}{2}}\put(4.5,-0.5){\pos{}{r}}
\put(-0.5,0.5){\spos{}{1}}\put(-0.5,1.5){\spos{}{2}}\put(-0.5,2.5){\spos{}{3}}
\put(-0.5,3.5){\pos{}{s}}
\put(2,4){\pos{}{p'=4}}
\end{picture}}

\nc{\gridfive}{\begin{picture}(7,6)(1,1)
\multiput(0,0)(0,1){5}{\line(1,0){6}}
\multiput(0,0)(2,0){4}{\line(0,1){4}}
\put(1.05,0.5){\spos{}{0}}\put(3,0.5){\spos{}{7/16}}\put(5,0.5){\spos{}{3/2}}
\put(1,1.5){\spos{}{1/10}}\put(3,1.5){\spos{}{3/80}}\put(5,1.5){\spos{}{3/5}}
\put(1,2.5){\spos{}{3/5}}\put(3,2.5){\spos{}{3/80}}\put(5,2.5){\spos{}{1/10}}
\put(1,3.5){\spos{}{3/2}}\put(3,3.5){\spos{}{7/16}}\put(5.05,3.5){\spos{}{0}}
\put(1.05,-0.5){\spos{}{1}}\put(3.05,-0.5){\spos{}{2}}
\put(5.05,-0.5){\spos{}{3}}\put(6.5,-0.5){\pos{}{r}}
\put(-0.5,0.5){\spos{}{1}}\put(-0.5,1.5){\spos{}{2}}\put(-0.5,2.5){\spos{}{3}}
\put(-0.5,3.5){\spos{}{4}}\put(-0.5,4.5){\pos{}{s}}
\put(3,5){\pos{}{p'=5}}
\end{picture}}

\nc{\gridsix}{\begin{picture}(8,7)(1,1)
\multiput(0,0)(0,1){6}{\line(1,0){8}}
\multiput(0,0)(2,0){5}{\line(0,1){5}}
\put(1.05,0.5){\spos{}{0}}\put(3,0.5){\spos{}{2/5}}\put(5,0.5){\spos{}{7/5}}
\put(7.05,0.5){\spos{}{3}}
\put(1,1.5){\spos{}{1/8}}\put(3,1.5){\spos{}{1/40}}\put(5,1.5){\spos{}{21/40}}
\put(7,1.5){\spos{}{13/8}}
\put(1,2.5){\spos{}{2/3}}\put(3,2.5){\spos{}{1/15}}\put(5,2.5){\spos{}{1/15}}
\put(7,2.5){\spos{}{2/3}}
\put(1,3.5){\spos{}{13/8}}\put(3,3.5){\spos{}{21/40}}\put(5,3.5){\spos{}{1/40}}
\put(7,3.5){\spos{}{1/8}}
\put(1.05,4.5){\spos{}{3}}\put(3,4.5){\spos{}{7/5}}\put(5,4.5){\spos{}{2/5}}
\put(7.05,4.5){\spos{}{0}}
\put(1.05,-0.5){\spos{}{1}}\put(3.05,-0.5){\spos{}{2}}
\put(5.05,-0.5){\spos{}{3}}\put(7.05,-0.5){\spos{}{4}}\put(8.5,-0.5){\pos{}{r}}
\put(-0.5,0.5){\spos{}{1}}\put(-0.5,1.5){\spos{}{2}}\put(-0.5,2.5){\spos{}{3}}
\put(-0.5,3.5){\spos{}{4}}\put(-0.5,4.5){\spos{}{5}}\put(-0.5,5.5){\pos{}{s}}
\put(4,6){\pos{}{p'=6}}
\end{picture}}

\nc{\gridtable}{\begin{picture}(12,13)
\put(3,1){\gridsix}\put(0,10){\gridfour}\put(8,9){\gridfive}
\end{picture}}

\begin{document}

\begin{center}
\title{LATTICE REALIZATIONS OF UNITARY MINIMAL\\
        MODULAR INVARIANT PARTITION FUNCTIONS}
\author{David L. O'Brien\footnote{E-mail: dlo@maths.mu.oz.au} and
        Paul A. Pearce\footnote{E-mail: pap@maths.mu.oz.au}}
\address{Mathematics Department, University of Melbourne,\\
            Parkville, Victoria 3052, Australia }

\begin{abstract}
\begin{quotation}
The conformal spectra of the critical dilute \ade lattice models are studied
numerically.
The results strongly indicate that, in branches 1 and 2, these models provide
realizations of the complete \ade classification of unitary minimal modular
invariant
partition functions given by Cappelli, Itzykson and Zuber. In branches 3
and 4 the results
indicate that the modular invariant partition functions factorize. Similar
factorization
results are also obtained for two-colour lattice models.
\end{quotation}
\end{abstract}
\end{center}

\section{Introduction}
\noindent
It is well established that the critical behaviour of two-dimensional
lattice models is described by conformal field theory or, to put it another
way,
that the continuum limit of critical lattice models provide realizations of
two-dimensional conformal field theories. An important class of conformal field
theories is the unitary minimal series with central charge $c<1$. In this case
a
complete \ade classification of the theories has been obtained by Cappelli,
Itzykson and Zuber~\cite{CIZ87}. In this paper we present compelling numerical
evidence to show that the critical dilute
\ade lattice models~\cite{WNS92,Roche92,WarnPSN94} provide realizations of this
complete \ade classification of unitary minimal conformal field theories, as
conjectured by Roche~\cite{Roche92}.  The layout of the paper is as follows. In
Section~2 we describe the minimal conformal field theories and their
\ade classification. In Section~3 we define the critical \ade models due to
Pasquier~\cite{Pasq87} and their dilute and
two-colour~\cite{WarnNien} generalizations. We also summarize the conjectured
modular invariant partition functions for these models. Finally, in Section~4,
we present the  numerical results that confirm the conjectured modular
invariant
partition functions.

\section{Conformal Field Theory}
\setcounter{equation}{0}

\subsection{Minimal Models}
In 1984 Belavin, Polyakov and Zamolodchikov~\cite{BPZ84} introduced the minimal
series of conformally invariant field theories.  These models are characterized
by a central charge $c<1$ which is restricted to the discrete values
\be
c=1-{6(p-p')^2\over pp'}
\ee
with $p$ and $p'$ coprime positive integers.
The conformal weights of the minimal series  are given by the Kac formula
\be
\Delta=\Delta_{r,s}^{(p,p')}={(rp'-sp)^2-(p'-p)^2\over 4pp'}
\ee
with
\be
1\le r\le p-1,\quad 1\le s\le p'-1.
\ee
Moreover, Friedan, Qiu and Shenker~\cite{FQS84} showed that if the theory is
unitary, then the central charge is further restricted by $|p-p'|=1$, and if in
fact $p'-p=1$,
\be
c=1-{6\over p'(p'-1)},\qquad p'=4,5,6,\ldots
\ee
The grids of conformal weights for $p'=4,5$ and $6$ are shown in Figure~1.
\begin{figure}[htbp]
\setlength{\unitlength}{6mm}
\begin{center}
\gridtable
\end{center}
\caption{Grids of conformal weights for the unitary minimal models
with $p'=4,5,6$ and $p=p'-1$. The table with $p'=4,\ c=1/2$ is identified with
the Ising model, $p'=5,\ c=7/10$ is identified with the tricritical Ising model
and $p'=6,\ c=4/5$ with the tetracritical
Ising model. The odd rows of the $p'=6$ Kac table give the critical
exponents of the
3-state Potts model.}
\end{figure}

\subsection{\ade Classification of Modular Invariant Partition Functions}

For a conformal field theory on a torus, modular invariance~\cite{Cardy}
implies further
constraints on the theory. The requirement of modular invariance is strong
enough to fix
the operator content. In fact, Cappelli, Itzykson and Zuber~\cite{CIZ87}
have obtained a
complete classification of minimal modular invariant partition functions.
Remarkably they
obtain two series in one-to-one correspondence with the \ade classical Lie
algebras, one labelled by $(A,G)$ and the other by $(G,A)$, with Coxeter
numbers $(p,p')$ in each case, and $p'>p$.  The \ade classification of minimal
modular invariant partition functions is shown in Table~1. The Virasoro
characters in this table are defined by
\be
\chi_{r,s}(q)
={q^{-c/24+\Delta_{r,s}^{(p,p')}}\over Q(q)}\sum_{n=-\infty}^\infty
\left\{q^{\textstyle{n(npp'+rp'-sp)}}
     - q^{\textstyle{(np'+s)(np+r)}}\right\}\qquad\mbox{}
\ee
where $q$ is the modular parameter and
\be  Q(q)=\disp{\prod_{n=1}^\infty (1-q^n)}.
\ee
\begin{table}[tbh]
\begin{center}
\small
\renewcommand{\arraystretch}{2.25}
\begin{tabular}{ll}   \hline $(A,G)$&Modular Invariant Partition Function
\\ \hline
$(A_{p-1},A_{p'-1})$
&\rule[-15pt]{0pt}{30pt}$\disp{Z=\half\sum_{r=1}^{p-1}\sum_{s=1}^{p'-1}|\chi
_{r,s}|^2}$\\
\parbox[t]{0.99in}{
$\disp{(A_{p-1},D_{2\rho+2})}$\par $\quad\scriptstyle{p'=4\rho+2\ge 6}$}
&\rule[-15pt]{0pt}{30pt}$\disp{Z=\half\sum_{r=1}^{p-1}
    \Biggl\{\sum_{s=1\atop s\ {\rm odd}}^{2\rho-1}
    |\chi_{r,s}+\chi_{r,4\rho+2-s}|^2+2|\chi_{r,2\rho+1}|^2\Biggr\}}$\\
\parbox[t]{0.99in}{
$\disp{(A_{p-1},D_{2\rho+1})}$\par $\quad\scriptstyle{p'=4\rho\ge 8}$}
&\rule[-15pt]{0pt}{30pt}$\disp{Z=\half\sum_{r=1}^{p-1}
    \Biggl\{\sum_{s=1\atop s\ {\rm odd}}^{4\rho-1}
    |\chi_{r,s}|^2\!+\!|\chi_{r,2\rho}|^2\!+\!
    \sum_{s=2\atop s\ {\rm even}}^{2\rho-2}

(\chi_{r,s}\bar{\chi}_{r,4\rho-s}+\bar{\chi}_{r,s}\chi_{r,4\rho-s})\Biggr\}}
$\\
\parbox[t]{0.99in}{
$(A_{p-1},E_6)$\par $\quad \scriptstyle{p'=12}$}
&\rule[-15pt]{0pt}{30pt}$\disp{Z=\half\sum_{r=1}^{p-1}
     \biggl\{|\chi_{r,1}+\chi_{r,7}|^2+|\chi_{r,4}+\chi_{r,8}|^2+
     |\chi_{r,5}+\chi_{r,11}|^2\biggr\}}$\\
\parbox[t]{0.99in}{
$(A_{p-1},E_7)$\par $\quad \scriptstyle{p'=18}$}
&\rule[-15pt]{0pt}{30pt}$\disp{Z=\half\sum_{r=1}^{p-1}
     \biggl\{|\chi_{r,1}+\chi_{r,17}|^2+|\chi_{r,5}+\chi_{r,13}|^2+
     |\chi_{r,7}+\chi_{r,11}|^2}$\\ &\qquad$\disp{\qquad\hbox{}+|\chi_{r,9}|^2
+[(\chi_{r,3}+\chi_{r,15})\bar{\chi}_{r,9}+
     (\bar{\chi}_{r,3}+\bar{\chi}_{r,15})\chi_{r,9}]\biggr\}}$\\
\parbox[t]{0.99in}{
$(A_{p-1},E_8)$\par $\quad \scriptstyle{p'=30}$}
&\rule[-15pt]{0pt}{30pt}$\disp{Z=\half\sum_{r=1}^{p-1}
     \biggl\{|\chi_{r,1}+\chi_{r,11}+\chi_{r,19}+\chi_{r,29}|^2}$\\

&\qquad$\disp{\qquad\hbox{}+|\chi_{r,7}+\chi_{r,13}+\chi_{r,17}+\chi_{r,23}|
^2
     \biggr\}}$ \vspace{4pt}\\ \hline
\end{tabular}
\renewcommand{\arraystretch}{1.0}
\normalsize
\end{center}
\caption{\ade classification of minimal modular invariant partition
functions. The
central charges are
$c=1-{6(p-p')^2\over pp'}$, $\chi_{r,s}=\chi_{r,s}(q)$ are Virasoro
characters and
bars denote complex conjugates. In this series $r,s$ are Coxeter exponents of
$(A,G)$. There is a second series where $r,s$ are Coxeter exponents of $(G,A)$.
In both series $p'>p$, and the unitary minimal models have $p'-p=1$.}
\label{MIPFs}
\end{table}

Here we are primarily interested in the unitary minimal models with
$p'-p=1$. In this case the two \ade series correspond to
\be
(A,G)=\cases{(A_{p'-2},A_{p'-1})&\cr (A_{p'-2},D_{(p'+2)/2})&\cr
(A_{10},E_6)&\cr
(A_{16},E_7)&\cr (A_{28},E_8)}\qquad\qquad (G,A) =\cases{(A_{p-1},A_{p})&\cr
(D_{(p+2)/2},A_{p})&\cr (E_6,A_{12})&\cr (E_7,A_{18})&\cr (E_8,A_{30})}
\ee
with central charges
\be
c=1-\frac{6}{p'(p'-1)}=1-\frac{6}{p(p+1)}
\ee
The Coxeter numbers and the Coxeter exponents of the classical
\ade Lie algebras are shown in Table~2. The Dynkin diagrams are shown in
Figure~2.  Some
members of these series are identified as follows:
\be
\begin{array}{ll} (A_2,A_3)=\mbox{critical Ising}&c=1/2\\
(A_4,D_4)=\mbox{critical
3-state Potts}&c=4/5\\
(A_3,A_4)=\mbox{tricritical Ising}&c=7/10\\
(D_4,A_6)=\mbox{tricritical 3-state Potts}&c=6/7\\
\end{array}
\ee
For this reason we will refer to the $(A,G)$ series as the critical series
and the
$(G,A)$ series as the tricritical series. In particular, the modular invariant
partition functions of the critical and tricritical 3-state Potts models are
\bea
(A_4,D_4):&&\quad Z=
\half \sum_{r=1}^4\left\{|\chi_{r,1}+\chi_{r,5}|^2+2|\chi_{r,3}|^2\right\}\quad
(p=5,p'=6)\\
(D_4,A_6):&&\quad Z=
\half\sum_{s=1}^6\left\{|\chi_{1,s}+\chi_{5,s}|^2+2|\chi_{3,s}|^2\right\}\quad
(p=6,p'=7).
\eea

\begin{table}[htbp]
\begin{center}
\begin{tabular}{ccc}
\hline \rule[-10pt]{0pt}{25pt}$G$&$h$&$s$\\
\hline
\rule[-10pt]{0pt}{25pt}$A_L$&$L+1$&$1,2,3,\ldots,L$\\
\rule[-10pt]{0pt}{20pt}$D_L$&$2L-2$&$L-1,1,3,5,\ldots,2L-3$\\
\rule[-10pt]{0pt}{20pt}$E_6$&$12$&$1,4,5,7,8,11$\\
\rule[-10pt]{0pt}{20pt}$E_7$&$18$&$1,5,7,9,11,13,17$\\
\rule[-10pt]{0pt}{20pt}$E_8$&$30$&$1,7,11,13,17,19,23,29$\\
\hline
\end{tabular}
\end{center}
\caption{The Coxeter number $h$ and Coxeter exponents $s$ of the classical
\ade Lie
algebras.}\label{CoxExp}
\end{table}

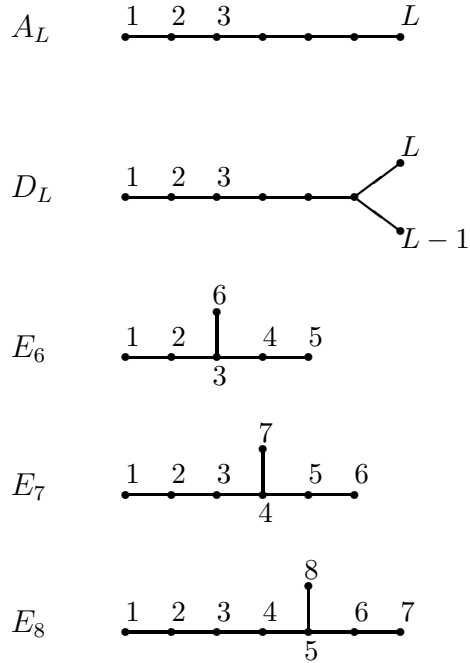
\begin{figure}[htbp]
\begin{center}
\begin{minipage}[t]{2.25in}

\setlength{\unitlength}{.012in}
\begin{center}
\begin{picture}(140,240)(-20,-160)

\put(-50,80){$A_L$}
\multiput(0,80)(20,0){6}{\line(1,0){20}}
\multiput(0,80)(20,0){7}{\circle*{4}}
\put(0,85){\tiny 1} \put(20,85){\tiny 2} \put(40,85){\tiny 3}
\put(120,85){\tiny $L$}

\put(-50,10){$D_L$}
\multiput(0,10)(20,0){5}{\line(1,0){20}}
\put(100,10){\line(4,3){20}} \put(100,10){\line(4,-3){20}}
\multiput(0,10)(20,0){6}{\circle*{4}}
\multiput(120,25)(0,-30){2}{\circle*{4}}
\put(0,15){\tiny 1} \put(20,15){\tiny 2} \put(40,15){\tiny 3}
\put(120,28){\tiny $L$}\put(120,-12){\tiny $L-1$}

\put(-50,-60){$E_6$}
\multiput(0,-60)(20,0){4}{\line(1,0){20}}
\multiput(0,-60)(20,0){5}{\circle*{4}}
\put(40,-40){\line(0,-1){20}} \put(40,-40){\circle*{4}}
\put(0,-55){\tiny 1} \put(20,-55){\tiny 2} \put(60,-55){\tiny 4}
\put(80,-55){\tiny 5}
\put(38,-72){\tiny 3} \put(38,-37){\tiny 6}

\put(-50,-120){$E_7$}
\multiput(0,-120)(20,0){5}{\line(1,0){20}}
\multiput(0,-120)(20,0){6}{\circle*{4}}
\put(60,-100){\line(0,-1){20}} \put(60,-100){\circle*{4}}
\put(0,-115){\tiny 1} \put(20,-115){\tiny 2} \put(40,-115){\tiny 3}
\put(80,-115){\tiny 5}
\put(100,-115){\tiny 6} \put(58,-132){\tiny 4} \put(58,-97){\tiny 7}

\put(-50,-180){$E_8$}
\multiput(0,-180)(20,0){6}{\line(1,0){20}}
\multiput(0,-180)(20,0){7}{\circle*{4}}
\put(80,-160){\line(0,-1){20}} \put(80,-160){\circle*{4}}
\put(0,-175){\tiny 1} \put(20,-175){\tiny 2} \put(40,-175){\tiny 3}
\put(60,-175){\tiny 4}
\put(100,-175){\tiny 6} \put(120,-175){\tiny 7} \put(78,-192){\tiny 5}
\put(78,-157){\tiny 8}

\end{picture}
\vspace{.3in}  
\end{center}
\end{minipage}
\end{center}
\caption{The Dynkin diagrams of the classical \ade Lie algebras. The \ade
graphs classify
all connected graphs whose associated adjacency matrices have eigenvalues
strictly less than 2. The
eigenvalues of the adjacency matrices are in fact given by
$2\cos(s\pi/h)$ where $s$ ranges over the Coxeter exponents.}
\setlength{\unitlength}{.01in}
\end{figure}

The \ade classification of unitary minimal conformal field theories gives
an exhaustive
list of theories with
$c<1$. In other words, this is a complete list of universality classes
giving all
possible critical behaviours for two-dimensional statistical systems with
$c<1$. A natural
question to ask is whether a solvable lattice model can be found as a
representative of
each universality class allowed by the \ade classification.

\section{\ade Lattice Models and Their Modular Invariant Partition Functions}
\setcounter{equation}{0}

\subsection{Pasquier's \ade Models}
\noindent
By a remarkable coincidence, in the same year that Belavin, Polyakov and
Zamolodchikov
introduced the minimal conformal field theories, Andrews, Baxter and
Forrester~\cite{ABF84} solved the first infinite hierarchy of lattice
models in the form
of restricted solid-on-solid (RSOS) models. The spins in these models take
values on the
$A_L$ Dynkin diagram and are subject to the constraint that the state of
adjacent spins
on the square lattice must be adjacent on the $A_L$ diagram.
Huse~\cite{Huse84} showed
that the critical behaviour of these
$L$ height RSOS models is precisely described by the unitary
minimal series. Moreover, it turns out that the modular invariant partition
functions of
the ABF RSOS models give the
$(A_{L-1},A_L)$ series with $L=3,4,5,\ldots$

The lattice realizations of this critical series of modular invariant partition
functions was completed in 1987 by Pasquier~\cite{Pasq87} who generalized
the ABF models
by constructing solvable lattice models whose states take values on the
\ade graphs. The $A_L$ models of Pasquier are just the critical ABF RSOS
models.
We note that, although the $A$ and $D$ models admit off-critical elliptic
extensions, the
exceptional $E$ models can only be solved at criticality.
The face weights of Pasquier's critical \ade models are given by
\be
\wt Wabcdu=\!\!\raisebox{.35\unitlength}{\sqface abcdu00}\ ={\sin (\lambda
-u) \over
\sin \lambda }\delta_{a,c}A_{a,b}A_{a,d}
 +{\sin u \over \sin \lambda } \sqrt{{S_a S_c \over S_b S_d}}
\delta_{b,d}A_{a,b}A_{b,c}
\label{eq:cADEface}
\ee where the spins $a,b,c,d$ take values on the given \ade graph. The
parameter $u$
is called the spectral parameter. In the branches of interest here the spectral
parameter lies in the interval $0<u<\lambda$. The adjacency matrices are
given by
\be A_{a,b}=\cases{1,&$a$, $b$ connected\cr
               0,&otherwise.}
\ee The nonnegative components $S_a$ of the Perron-Frobenius eigenvector are
determined by
\be
\sum_b A_{a,b}S_b=2\cos\lambda\; S_a
\ee where $2\cos\lambda$ is the largest eigenvalue of the adjacency matrix and
\be
\lambda=\pi/h
\ee is called the crossing parameter. The Coxeter number $h$ is given in
Table~2.

Pasquier's \ade models include some much studied models in statistical
mechanics.
Some prototypes are shown in Figure~3. The modular invariant partition
functions of
Pasquier's critical \ade models precisely realize the $(A,G)$ series of
Cappelli,
Itzykson and Zuber. However, for many years
realizations of the $(G,A)$ series were missing.
\begin{figure}[htbp]
\setlength{\unitlength}{.014in}
$$
\begin{array}{rcccl} A_3&=&
\begin{picture}(50,15)(0,80)
\raisebox{2\unitlength} {\multiput(0,80)(20,0){2}{\line(1,0){20}}
\multiput(0,80)(20,0){3}{\circle*{4}}
\put(-2,85){\tiny 1} \put(18,85){\tiny 2} \put(38,85){\tiny 3} }
\end{picture}
&\ =\ &\mbox{Critical Ising}\\ \\  A_4&=&
\begin{picture}(70,15)(0,80)
\raisebox{2\unitlength}  {\multiput(0,80)(20,0){3}{\line(1,0){20}}
\multiput(0,80)(20,0){4}{\circle*{4}}
\put(-2,85){\tiny 1} \put(18,85){\tiny 2} \put(38,85){\tiny 3}
\put(58,85){\tiny 4} }
\end{picture}
&\ =\ &
\mbox{Tricritical Hard Squares}\\ \\  D_4&=&
\raisebox{-18\unitlength}{
\begin{picture}(50,45)(0,80)
\raisebox{20\unitlength} {
\put(0,80){\line(1,0){20}}\put(20,80){\line(1,1){15}}\put(20,80){\line(1,-1)
{15}}
\multiput(0,80)(20,0){2}{\circle*{4}}\put(35,65){\circle*{4}}
\put(35,95){\circle*{4}}
\put(-2,85){\tiny 1} \put(18,85){\tiny 2} \put(40,94){\tiny 4}
\put(40,62){\tiny 3}}
\end{picture}}
&\ =\ &\mbox{Critical 3-State Potts}
\end{array}
$$
\caption{Some prototype classical \ade lattice models.}
\end{figure}

\ignore{ {\large Affine:}
\begin{eqnarray*}  A_3^{(1)}&=&\hspace{.4in}
\raisebox{-22\unitlength}{
\begin{picture}(40,45)(0,80)
\raisebox{35\unitlength} {
\put(0,80){\line(1,0){30}}\put(0,80){\line(2,-3){15}}
\put(30,80){\line(-2,-3){15}}
\multiput(0,80)(30,0){2}{\circle*{4}}\put(15,58){\circle*{4}}
\put(-2,85){\tiny 1} \put(28,85){\tiny 2} \put(13,45){\tiny 3}}
\end{picture}}
\hspace{.4in} =\ \mbox{3-Coloring Problem}\\ \\  D_4^{(1)}&=&\hspace{.3in}
\raisebox{-18\unitlength}{
\begin{picture}(50,45)(0,80)
\raisebox{20\unitlength} {
\put(20,80){\line(-1,1){15}}\put(20,80){\line(-1,-1){15}}
\put(20,80){\line(1,1){15}}\put(20,80){\line(1,-1){15}}
\put(5,65){\circle*{4}}\put(5,95){\circle*{4}}
\put(20,80){\circle*{4}}\put(35,65){\circle*{4}}\put(35,95){\circle*{4}}
\put(-5,62){\tiny 1} \put(-5,94){\tiny 2} \put(18,85){\tiny 3}
\put(40,94){\tiny 4} \put(40,62){\tiny 5}}
\end{picture}}
\hspace{.3in} =\ \mbox{4-State Potts}\\ \\  D_5^{(1)}&=&
\raisebox{-18\unitlength}{
\begin{picture}(70,45)(0,80)
\raisebox{20\unitlength} {
\put(20,80){\line(-1,1){15}}\put(20,80){\line(-1,-1){15}}\put(20,80){\line(1
,0){20}}
\put(40,80){\line(1,1){15}}\put(40,80){\line(1,-1){15}}
\put(5,65){\circle*{4}}\put(5,95){\circle*{4}}
\multiput(20,80)(20,0){2}{\circle*{4}}\put(55,65){\circle*{4}}
\put(55,95){\circle*{4}}
\put(-5,62){\tiny 1} \put(-5,94){\tiny 2} \put(18,85){\tiny 3}
\put(38,85){\tiny 4}
\put(60,94){\tiny 5} \put(60,62){\tiny 6}}
\end{picture}}
\ =\ \mbox{Magnetic Hard Squares}
\end{eqnarray*} }
\goodbreak

\subsection{Dilute \ade Models}
\noindent
In 1992 Warnaar, Nienhuis and Seaton~\cite{WNS92} and Roche~\cite{Roche92}
independently obtained a second series of solvable lattice models whose
states take
values on the \ade graphs. These lattice models are called the dilute \ade
models.
The face weights of the dilute \ade lattice models at criticality are given by
\bea  &&\ \ \wt Wabcdu\ =\ \rho_1 (u)\delta_{a,b,c,d} +\rho_2 (u)
\delta_{a,b,c}
A_{a,d} + \rho_3 (u)\delta_{a,c,d} A_{a,b}\non\\   &&\mbox{}+\sqrt{S_a \over
S_b}\rho_4 (u) \delta_{b,c,d} A_{a,b}
     +\sqrt{S_c \over S_a}\rho_5 (u) \delta_{a,b,d} A_{a,c}+
\rho_6 (u) \delta_{a,b} \delta_{c,d} A_{a,c}\\ &&\mbox{}+\rho_7 (u)
\delta_{a,d}
\delta_{c,b} A_{a,b}+
\rho_8(u) \delta_{a,c}A_{a,b} A_{a,d}+
       \sqrt{{S_a S_c \over S_b S_d}} \rho_9 (u) \delta_{b,d}
A_{a,b}A_{b,c}\non
\eea where, as before, the adjacency matrix is
\be  A_{a,b}=\cases{1,&$a, b$ adjacent\cr 0,&otherwise}
\ee and the Perron-Frobenius vector is given by
\be
\sum_b A_{a,b}\,S_b=2\cos\left({\pi\over h}\right)\,S_a.
\ee The effective adjacency graph is given by adding a loop to each node of
the \ade
graphs, that is, the spin states at adjacent sites of the lattice are either
the
same or adjacent on the \ade graph.  The generalized Kronecker delta is
\be
\delta_{a,b,c,\ldots}=\cases{1,&$a=b=c=\ldots$\cr 0,&otherwise}
\ee and the trigonometric weight functions are
\bea
\rho_1 (u)&=&1\ +\ {\sin u
\sin (3\olambda-u) \over \sin (2\olambda )\sin (3\olambda )}\non\\
\rho_2 (u)&=&\rho_3 (u)\ =\ \ {\sin (3\olambda-u)\over \sin(3\olambda )} \non\\
\rho_4 (u)&=&\rho_5 (u)\ =\ \ \epsilon{\sin u \over \sin(3\olambda )}\non\\
\rho_6 (u)&=&\rho_7 (u)\ =\ \ \epsilon{\sin u \sin (3\olambda-u) \over
\sin(2\olambda )\sin (3\olambda )} \\
\rho_8 (u)&=&{\sin (2\olambda-u) \sin (3\olambda-u) \over \sin (2\olambda
)\sin(3\olambda)}\non\\
\rho_9 (u)&=&-\;{\sin u \sin (\olambda-u) \over \sin (2\olambda)\sin (3\olambda
)}.\non
\eea
Here $\epsilon=\pm1$; the choices $\epsilon=1$ in the $u>0$ branches and
$\epsilon=-1$ in the $u<0$ branches ensure positive Boltzmann weights at the
isotropic points $u=\gamma/2$.

The dilute \ade models are solvable for two choices of $\lambda$
\be
   \olambda=\cases{\displaystyle{(h-1)\pi\over 4h},&branches 1 and 4\cr\cr
                   \displaystyle{(h+1)\pi\over 4h},&branches 2 and 3.}
\ee
The physical branches are summarized in Table~3.
The central charges of these models in branches 1 and 2 are given by
\cite{WarnBN92,WNS92,WarnPSN94,PearZ95}
\be
   c=\cases{1-\displaystyle{6\over h(h+1)},&branch 1\cr\cr
         1-\displaystyle{6\over h(h-1)},&branch 2.}
\ee
This suggests identifying the universality classes of the first few dilute \ade
models as follows:
\be
\begin{array}{ll}
\mbox{branch 2:}\quad A_3=\mbox{critical Ising}&c=1/2\\
\mbox{branch 1:}\quad A_3=\mbox{tricritical Ising}&c=7/10\\
\mbox{branch 2:}\quad D_4=\mbox{critical 3-state Potts}&c=4/5\\
\mbox{branch 1:}\quad D_4=\mbox{tricritical 3-state Potts}&c=6/7\\
\end{array}
\ee Notice that the dilute $A_3$ and $D_4$ are not the usual Ising and
3-state Potts
models, they just have the same $\mbox{\boldmath{$Z_2$}}$ and
$\mbox{\boldmath{$Z_3$}}$ symmetries
and lie in the
same universality classes.


\nc{\round}[1]{\left(#1\right)}
\nc{\lam}{\lambda}

\nc{\pointy}[1]{\left<#1\right>}
\nc{\curly}[1]{\left\{#1\right\}}
\nc{\abs}[1]{\left|#1\right|}
\nc{\lsum}[3]{\sum_{#1=#2}^{#3}}
\nc{\lprod}[3]{\prod_{#1=#2}^{#3}}
\nc{\refeq}[1]{(\protect\ref{#1})}


\nc{\Lam}{\Lambda}
\nc{\Om}{\Omega}
\nc{\gam}{\gamma}


\nc{\embolden}[1]{\mbox{\boldmath ${#1}$}}
\nc{\Zset}{\embolden{Z}}

\nc{\Order}[1]{{\rm O}\round{#1}}
\nc{\order}[1]{{\rm o}\round{#1}}
\nc{\q}[2]{q^{\frac{#1}{#2}}}
\nc{\mq}[3]{#1\,\q{#1}{#2}}
\nc{\square}[1]{\left[#1\right]}

\begin{table}[htbp]
\begin{center}
\begin{tabular}{|l||l|l|l|l|} \hline
         & Crossing param. & Inversion pt & Phys.\ region &
                                                            Central charge
\\ \hline
Branch 1 & $\lam=\frac{\pi}{4}\round{1-\frac{1}{h}}$ & $\gam=3\lam$ &
               $u\in\round{0,\gam}$ & $c=1-\frac{6}{h\round{h+1}}$ \\ Branch 2
&
$\lam=\frac{\pi}{4}\round{1+\frac{1}{h}}$ & $\gam=3\lam$ &
               $u\in\round{0,\gam}$ & $c=1-\frac{6}{h\round{h-1}}$ \\ Branch 3
&
$\lam=\frac{\pi}{4}\round{1+\frac{1}{h}}$ & $\gam=3\lam-\pi$ &
               $u\in\round{\gam,0}$ &
$c=\frac{3}{2}-\frac{6}{h\round{h+1}}$ \\ Branch 4
& $\lam=\frac{\pi}{4}\round{1-\frac{1}{h}}$ & $\gam=3\lam-\pi$ &
               $u\in\round{\gam,0}$ & $c=\frac{3}{2}-\frac{6}{h\round{h-1}}$ \\
\hline
\end{tabular}
\end{center}
\caption{Physical branches and central charges of the dilute \ade lattice
models}
\label{dilconst}
\end{table}

The dilute \ade lattice models in branch 2 in fact give a second
realization of the
$(A,G)$ series of Cappelli, Itzykson and Zuber. More importantly, as we show
in the next section, the dilute \ade lattice models in branch 1
precisely realize the missing $(G,A)$ series. The dilute \ade models thus
give a
complete realization of all unitary minimal conformal field theories.

\subsection{Two-Colour \ade Models}

The two-colour models, obtained by Warnaar and Nienhuis \cite{WarnNien}, are
dense RSOS models built on pairs of \ade adjacency graphs.  Each site on the
lattice carries two heights, one from each graph.  In moving between adjacent
sites, one of the heights remains constant and the other varies as permitted by
its corresponding adjacency graph.  If $G^1$ and $G^2$ are the adjacency
matrices of the two graphs, the effective adjacency matrix of the two-colour
model is therefore
$A=A^1 + A^2$,
where
\be
A^1= G^1 \otimes I,\qquad
A^2 = I \otimes G^2
\ee
With a slight abuse of notation we will denote such a model by $G^1\otimes
G^2$.  The Perron-Frobenius vector of $A$ is the tensor product
$S=S^1\otimes  S^2$
of the  Perron-Frobenius vectors of the underlying graphs.
In constructing the two-colour models it is assumed that the graphs $G^1$
and $G^2$ have
the same largest eigenvalue and hence the same Coxeter numbers.

\begin{table}[htbp]
\begin{center}
\begin{tabular}{|l||l|l|l|l|} \hline
         & Crossing param. & Inversion pt & Phys.\ region &
                                                            Central charge
\\ \hline
Branch 1 & $\lam=\frac{\pi}{2}\round{1-\frac{1}{h}}$ & $\gam=3\lam-\pi$ &
               $u\in\round{0,\gam}$ & $c=2\round{1-\frac{6}{h\round{h+1}}}$
\\ Branch 2 &
$\lam=\frac{\pi}{2}\round{1-\frac{1}{h}}$ & $\gam=3\lam-2\pi$ &
               $u\in\round{\gam,0}$ & $c=2\round{1-\frac{6}{h\round{h-1}}}$ \\
\hline
\end{tabular}
\end{center}
\caption{Physical branches and central charges of the two-colour \ade
lattice models}
\label{tcconst}
\end{table}

Explicitly, the state at each site
is given by an ordered pair $a=(a_1,a_2)$ so that
\bea
S_a &=& S_{a_1}^1 S_{a_2}^2 \\
A_{a,b}^i &=& G_{a_i,b_i}^i \delta_{a_{3-i},b_{3-i}}.
\eea
In terms of these entities, the face weights of the critical lattice model
are given by
\bea
\wt{W}{a}{b}{c}{d}{u}&=&\sum_{i=1}^2\biggl[\rho_1(u)A_{a,b}^i A_{b,c}^{3-i}
A_{c,d}^i A_{d,a}^{3-i}\biggr.\nonumber\\
& &\mbox{}+\delta_{a,c}\round{\rho_2(u)A_{a,b}^i
A_{a,d}^i+\rho_3(u)A_{a,b}^i A_{a,d}^{3-i}}\\
& &\mbox{}+\biggl. \sqrt{S_a S_c\over S_b S_d}\;
\delta_{b,d}\round{\rho_4(u)
A_{a,b}^i A_{b,c}^{3-i}+\rho_5(u)A_{a,b}^{3-i}A_{b,c}^i}
\biggr]\nonumber
\eea
where
\bea
\rho_1(u)&=&\epsilon\,\frac{\sin u\sin(3\lam-u)}{\sin\lam\sin3\lam} \non\\
\rho_2(u)&=&\frac{\sin(\lam-u)\sin(3\lam-u)}{\sin\lam\sin3\lam} \non\\
\rho_3(u)&=&\frac{\sin(3\lam-u)}{\sin3\lam} \\
\rho_4(u)&=&-\frac{\sin u\sin(2\lam-u)}{\sin\lam\sin3\lam} \non\\
\rho_5(u)&=&-\epsilon\,\frac{\sin u}{\sin3\lam}\non
\eea
The two-colour models have two physical regimes as summarized in
Table~\ref{tcconst} in terms of the Coxeter number $h$ of the underlying
graphs.
The choice of the sign factor $\epsilon=\pm 1$ such that $\epsilon=1$ in branch
1 and $\epsilon=-1$ in branch 2 ensures that the Boltzmann weights are positive
at the isotropic points $u=\gam/2$.

\subsection{Conjectured Modular Invariant Partition Functions}

The partition function of a critical lattice model on a finite
$\ell\times\ell'$ periodic lattice or torus can be
written as
\be
Z_{\ell,\ell'}\sim\exp(-\ell\ell' f)Z(q)
\ee
where $f$ is the bulk free energy and $Z(q)$ is a universal term describing the
leading finite-size corrections in the limit of $\ell, \ell'$ large with the
aspect ratio $\d=\ell'/\ell$ fixed. The argument $q$ is the modular parameter.
For a spatially isotropic model, it is simply related to the aspect ratio
$\d$ by $q=\exp(-2\pi\d)$.

The modular invariant partition functions of the dilute \ade models built on
the classical graph $G$ with Coxeter number $h$ are conjectured to be
as follows:
\be
\begin{array}{ll}
\mbox{Branch 1:} & (G,A_h) \\
\mbox{Branch 2:} & (A_{h-2},G) \\
\mbox{Branch 3:} & (G,A_h)\times (A_2,A_3) \\
\mbox{Branch 4:} & (A_{h-2},G)\times (A_2,A_3).
\end{array}
\label{conjecture}
\ee
Here the modular parameter is
\be
q=\exp(2\pi i\tau),\qquad \tau={\ell'\over\ell}\exp[i(\pi-\theta)]
\ee
and the effective angle $\theta$ \cite{KimPearce} is given by
\be
\theta=\cases{\disp{\pi u\over 3\lambda},&branches 1 and 2\cr\cr
              \disp{\pi u\over 3\lambda-\pi},&branches 3 and 4.}
\ee

The modular invariant partition functions of the two-colour \ade models are
conjectured to be as follows, where once again $h$ is the Coxeter number of
the underlying graphs $G^1$ and $G^2$:
$$
\begin{array}{ll}
\mbox{Branch 1:} & (G^1,A_h)\times (G^2,A_h) \\
\mbox{Branch 2:} & (A_{h-2},G^1)\times (A_{h-2},G^2),
\end{array}
$$
Here the modular parameter $q$ is as before but now the effective angle
is
\be
\theta=\cases{\disp{\pi u\over 3\lambda-\pi},&branch 1\cr\cr
              \disp{\pi u\over 3\lambda-2\pi},&branch 2.}
\ee

\section{Numerical Results}
\setcounter{equation}{0}

The central charges and scaling dimensions of critical lattice models can
be extracted
\cite{Affleck} from the finite-size corrections to the eigenvalues of the row
transfer matrices
\be
\langle \mbox{\boldmath $a$}|\mbox{\boldmath $T$}(u)|\mbox{\boldmath
$b$}\rangle
=\prod_{j=1}^N \wt W{a_j}{a_{j+1}}{b_{j+1}}{b_j}u.
\ee
Specifically, the finite-size corrections to the largest
eigenvalue $\Lam_0$ of a periodic transfer matrix with $N$ faces take the form
\be
\frac{1}{N} \log\Lam_0(u) = -f(u)+\frac{\pi c}{6N^2}\sin\theta(u)
+
\order{\frac{1}{N^2}}, \label{fsize}
\ee
where  $f$ is the free energy, $c$ is the central charge and $\theta(u)$ is
the effective
angle as defined in Section~3.4. At an isotropic point for a square ordered
phase $u$ is
fixed such that $\theta=\pi/2$.  The finite-size corrections to the
next-largest
eigenvalues $\Lam_n$ with $n=1,2,3,\ldots$ take the form
\be
\frac{1}{N} \log\Lam_n(u) =-f(u)+\frac{2\pi}{N^2} \square
{\round{\frac{c}{12}-x_n}
  \sin\theta(u)-is_n\cos\theta(u)} + \order{\frac{1}{N^2}}, \label{fdim}
\ee
where $x_n=\Delta+\overline{\Delta}$ and $s_n=\Delta-\overline{\Delta}$ are
respectively
the scaling dimension and spin. The scaling dimension takes fractional
values whereas
the spin is restricted to integer values.

The free energies of the dilute and two-colour models are calculated by solving
the appropriate inversion relations
\be
\kappa(u)\kappa(-u)=\rho(u)\rho(-u),\qquad
\kappa(u)=\kappa(\gamma-u)
\ee
where $f(u)=-\log\kappa(u)$ is the free energy and
\be
\rho(u)=\cases{\disp{\sin(2\lambda-u)\sin(3\lambda-u)\over\sin2\lambda\sin3
\lambda},
&dilute models\cr \cr
\disp{\sin(\lambda-u)\sin(3\lambda-u)\over\sin\lambda\sin3\lambda},&
two-colour models.}
\ee
The free energy of the critical dilute models is
given by \cite{WarnPSN94}
\be
f(u) = -2\int_{-\infty}^{\infty}
        \frac{\cosh(\pi-5\lam)x\cosh\lam x\sinh(\gam-u)x\sinh ux}
      {x\sinh\pi x\cosh\gam x}\,dx
\ee
and the free energy of the critical two-colour models is given by
\be
f(u) = -2\int_{-\infty}^{\infty}
 \frac{\cosh(\pi-2\lam)x\cosh2(\pi-2\lam)x\sinh(\gam-u)x\sinh ux}
      {x\sinh\pi x\cosh\gam x}\,dx.
\ee
In these expressions $\gam$ is the inversion point in the appropriate
branch.  The dilute and two-colour models have respectively four and two
critical
branches.  Tables~\ref{dilconst} and \ref{tcconst} summarize the crossing
parameters,
inversion points and central charges in each of the physical branches in
terms of the
Coxeter numbers $h$ of the underlying graphs as given in Table~\ref{CoxExp}.

Given the free energy, the central charge $c$ is estimated by calculating a
sequence of largest eigenvalues $\Lam_0$ for increasing values of $N$ and
applying a
suitable extrapolation scheme.  Once the central charge is determined,
\eqn{fdim} then
allows estimation of the scaling dimensions by a similar procedure using
the next-largest
eigenvalues $\Lam_n$. To obtain accurate values for the central charges and
scaling
dimensions we need to calculate eigenvalues for $N$ as large as possible.  It
is therefore convenient to prediagonalize the transfer matrices into block
diagonal form
using the eigenvectors of the shift operator $\Om=\mbox{\boldmath $T$}(0)$
and, in
the case of the $A$ and $D$ models, the reflection operator
$\mbox{\boldmath$R$}$ which
arises from the
$\mbox{\boldmath{$Z_2$}}$ symmetry of the $A$ and $D$ Dynkin diagrams.  Taken
together,
these operators
reduce the transfer matrices to $2N$ diagonal blocks.

Once a sequence of eigenvalues for increasing $N$ is obtained, equations
\eqn{fsize} and \eqn{fdim} imply that, for large $N$, the graph of
$(\log\Lam_n/N+f)$ against $1/N^2$ should approximate a line through the
origin.
However, since the $\order{N^{-2}}$ corrections tend to vanish fairly
slowly, a parabolic
fit gives better results.  A simple extrapolation scheme to extract the
$1/N$ term is to
discard all but the last two eigenvalues in the sequence and take the linear
coefficient of the parabola passing through these two points and the origin.
We have performed this calculation to find numerically the central charges
of a variety of
dilute and two-colour models as well as the scaling dimensions of the
dilute $A_3$,
$A_4$, $D_4$, and two-colour $A_4\otimes A_4$ models.  The approximate
central charges
are summarized in Tables~\ref{dilcent12} to \ref{tccent} and the
approximate scaling
dimensions are summarized in Tables~\ref{sdA312} to \ref{sdA4A412}.

\begin{table}[t]
\begin{center}
\begin{tabular}{|l||l|c|l||l|c|l||l|}   \hline
      & \multicolumn{3}{c||}{Branch 1} & \multicolumn{3}{c||}{Branch 2}
      &  \\ \cline{2-4} \cline{5-7} Model & Approx. &
\multicolumn{2}{c||}{Exact} &
        Approx. & \multicolumn{2}{c||}{Exact} & $N_{\rm max}$
\\ \hline
$A_3$   & 0.699999 & 7/10 & 0.7 & 0.500000 & 1/2 & 0.5 & 12 \\
$A_4$ & 0.799997 & 4/5 & 0.8 & 0.699997 & 7/10 & 0.7 & 10 \\
$A_5$,$D_4$ & 0.857140 & 6/7 & 0.857143\ldots & 0.799997 & 4/5 & 0.8 & 10 \\
$A_7$,$D_5$ & 0.916660 & 11/12 & 0.916666\ldots &
                                     0.892849 & 25/28 & 0.892857\ldots & 9\\
$A_{11}$,$D_7$,$E_6$
          & 0.961531 & 25/26 & 0.961538\dots &
                   0.954536 & 21/22 & 0.954545\ldots & 9 \\
$A_{17}$,$D_{10}$,$E_7$
          & 0.982448 & 56/57 & 0.982456\ldots &
                   0.980383 & 50/51 & 0.980392\ldots & 9 \\ \hline
\end{tabular}
\end{center}
\caption{Central charges of the dilute \ade lattice models in the $u>0$
branches.  The numerical approximations are in excellent agreement with the
exact values \protect\cite{WNS92} summarized in
Table~\protect\ref{dilconst}.}
\label{dilcent12}
\end{table}
\begin{table}[t]
\begin{center}
\begin{tabular}{|l||l|c|l||l|c|l||l|} \hline
      & \multicolumn{3}{c||}{Branch 3} & \multicolumn{3}{c||}{Branch 4}
      &  \\ \cline{2-4} \cline{5-7} Model & Approx. &
\multicolumn{2}{c||}{Exact} &
Approx. &
             \multicolumn{2}{c||}{Exact} & $N_{\rm max}$
\\ \hline
$A_3$   & 1.232 & 6/5 & 1.2 & 1.000 & 1 & 1 & 12 \\
$A_4$ & 1.327 & 13/10 & 1.3 & 1.199 & 6/5 & 1.2 & 10 \\
$A_5$,$D_4$ & 1.377 & 19/14 & 1.357\ldots & 1.299 & 13/10 & 1.3 & 10 \\
$A_7$,$D_5$ & 1.432 & 17/12 & 1.417\dots & 1.391 & 39/28 & 1.393\ldots & 8\\
$A_{11}$,$D_7$,$E_6$
          & 1.470 & 19/13 & 1.462\ldots & 1.453 & 16/11 & 1.455\ldots & 8 \\
$A_{17}$,$D_{10}$,$E_7$ & 1.487 & 169/114 & 1.482\ldots
          & 1.479 &151/102 & 1.480\ldots & 8 \\ \hline
\end{tabular}
\end{center}
\caption{Central charges of the dilute \ade lattice models in the $u<0$
branches.}
\label{dilcent34}
\end{table}

\begin{table}[hb]
\begin{center}
\begin{tabular}{|l||l|c|l||l|c|l||l|} \hline
      & \multicolumn{3}{c||}{Branch 1} & \multicolumn{3}{c||}{Branch 2}
      &  \\ \cline{2-4} \cline{5-7} Model & Approx. &
\multicolumn{2}{c||}{Exact} &
Approx. &
     \multicolumn{2}{c||}{Exact} & $N_{\rm max}$\\ \hline
$A_3 \otimes A_3$ & 1.423 & 7/5 & 1.4 & 0.9991 & 1 & 1 & 8 \\
$A_4 \otimes A_4$ & 1.606 & 8/5 & 1.6 & 1.3982 & 7/5 & 1.4 & 8 \\
$[A_5,D_4]\otimes[A_5,D_4]$
        & 1.712 & 12/7 & 1.714\ldots & 1.5930 & 8/5 & 1.6 & 6 \\
$[A_{11},D_7,E_6]\otimes$
        & 1.914 & 25/13 & 1.923\ldots & 1.8993 & 21/11 & 1.909\ldots & 6 \\
$\mbox{\ \ \ \ \ \ \ \ }[A_{11},D_7,E_6]$ & & & & & & & \\
\hline
\end{tabular}
\end{center}
\caption{Central charges of the two-colour \ade lattice models. Here
the notation
$[G,G']$ means either $G$ or $G'$.  The approximations match well the
predictions summarized in Table~\protect\ref{tcconst}.}
\label{tccent}
\end{table}

\begin{table}[htbp]
\begin{center}
\begin{tabular}{|l|c|l|c||l|c|l|c|} \hline
\multicolumn{4}{|c||}{Branch 1} & \multicolumn{4}{c|}{Branch 2} \\ \hline
Approx. &
\multicolumn{2}{c|}{Exact} & Mult. & Approx. &
\multicolumn{2}{c|}{Exact} & Mult. \\ \hline 0.0749999 & 3/40 & 0.075 & 1 &
0.125999 &
1/8 & 0.125 & 1 \\ 0.200000 & 1/5 & 0.2 & 1  & 0.998457 & 1 & 1 & 1 \\
0.874980 & 7/8 &
0.875 & 1 & 1.12489 & 9/8 & 1.125 & 2 \\ 1.07505 & 43/40 & 1.075 & 2 &
2.01084 & 2 & 2 &
2 \\ 1.20003 & 6/5 & 1.2 & 2 & 1.98828 & 2 & 2 & 2 \\ 1.19994 & 6/5 & 1.2 &
1 & 2.13060 &
17/8 & 2.125 & 2 \\ 1.87541 & 15/8 & 1.875 & 2 & & & & \\
\hline
\end{tabular}
\end{center}
\caption{Scaling dimensions and multiplicities for the dilute $A_3$ model in
the
$u>0$ branches.  We expect branch 1 to correspond to the partition function
labelled by $(A_3,A_4)$ and branch 2 to that labelled by $(A_2,A_3)$.  These
correspondences may be verified by comparison of the above approximations to
the exact exponents which arise in the partition function expansions of
\refeq{normexp}.}
\label{sdA312}
\end{table}
\begin{table}[htbp]
\begin{center}
\begin{tabular}{|l|c|l|c||l|c|l|c|} \hline
\multicolumn{4}{|c||}{Branch 3} & \multicolumn{4}{c|}{Branch 4} \\ \hline
Approx. &
\multicolumn{2}{c|}{Exact} & Mult. & Approx. &
\multicolumn{2}{c|}{Exact} & Mult. \\ \hline 0.0778 & 3/40 & 0.075 & 1 &
0.1250 & 1/8 &
0.125 & 1 \\ 0.1270 & 1/8 & 0.125 & 1 & 0.1250 & 1/8 & 0.125 & 1 \\ 0.2033
& 1/5 & 0.2 &
1 & 0.2500 & 1/4 & 0.25 & 1 \\ 0.2401 & 1/5 & 0.2 & 1 & 1.0000 & 1 & 1 & 1
\\ & & & &
1.0000 & 1 & 1 & 1 \\ & & & & 1.1250 & 9/8 & 1.125 & 1 \\ & & & & 1.1246 &
9/8 & 1.125 &
2 \\ & & & & 1.1248 & 9/8 & 1.125 & 1 \\ & & & & 1.1210 & 9/8 & 1.125 & 2
\\ & & & &
1.2450 & 5/4 & 1.25 & 2 \\ & & & & 1.2424 & 5/4 & 1.25 & 2 \\ & & & &
1.9987 & 2 & 2 & 2
\\ & & & & 1.9964 & 2 & 2 & 2 \\ & & & & 1.9949 & 2 & 2 & 1 \\
\hline
\end{tabular}
\end{center}
\caption{Scaling dimensions and multiplicities for the dilute $A_3$ model
in the $u<0$ branches.  Comparison with the expansions \refeq{prodexp} shows
branch 4 to be in excellent agreement with the partition function product
$(A_2,A_3)\times(A_2,A_3)$, and branch 3 to be in reasonable agreement with
$(A_3,A_4)\times(A_2,A_3)$.}
\label{sdA334}
\end{table}

\begin{table}[htbp]
\begin{center}
\begin{tabular}{|l|c|l|c||l|c|l|c|} \hline
\multicolumn{4}{|c||}{Branch 1} & \multicolumn{4}{c|}{Branch 2} \\ \hline
Approx. &
\multicolumn{2}{c|}{Exact} & Mult. & Approx. &
\multicolumn{2}{c|}{Exact} & Mult. \\
 \hline 0.050000 & 1/20 & 0.05 & 1 & 0.074999 & 3/40 & 0.075 & 1 \\
0.013333 & 2/15 &
0.133\ldots & 1 & 0.199997 & 1/5 & 0.2 & 1 \\ 0.250000 & 1/4 & 0.25 & 1  &
0.873958 & 7/8
& 0.875 & 1 \\ 0.799969 & 4/5 & 0.8 & 1  & 1.075250 & 43/40 & 1.075 & 2 \\
1.050060 &
21/20 & 1.05 & 2& 1.199460 & 6/5 & 1.2 & 2 \\ 1.049920 & 21/20 & 1.05 & 1&
1.196920 & 6/5
& 1.2 & 1 \\ 1.133360 & 17/15 & 1.133\ldots & 2& & & & \\
\hline
\end{tabular}
\end{center}
\caption{Scaling dimensions and multiplicities for the dilute $A_4$ model
in the $u>0$ branches.  Branch 1 agrees well with the expansion
\refeq{normexp} of the partition function $(A_4,A_5)$, as does branch 2 with
the
expansion of
$(A_3,A_4)$.}
\label{sdA412}
\end{table}
\begin{table}[htbp]
\begin{center}
\begin{tabular}{|l|c|l|c||l|c|l|c|} \hline
\multicolumn{4}{|c||}{Branch 3} & \multicolumn{4}{c|}{Branch 4} \\ \hline
Approx. &
\multicolumn{2}{c|}{Exact} & Mult. & Approx. &
\multicolumn{2}{c|}{Exact} & Mult. \\ \hline 0.0501 & 1/20 & 0.05 & 1 &
0.07500 & 3/40 &
0.075 & 1 \\ 0.1256 & 1/8 & 0.125 & 1 & 0.12500 & 1/8 & 0.125 & 1 \\ 0.1357
& 2/15 &
0.133\ldots & 1 & 0.19985 & 1/5 & 0.2 & 1 \\ 0.1918 & 7/40 & 0.175 & 1 &
0.20000 & 1/5 &
0.2 & 1 \\ 0.2498 & 1/4 & 0.25 & 1 & 0.32500 & 13/40 & 0.325 & 1 \\ & & & &
0.87506 & 7/8
& 0.875 & 1 \\ & & & & 0.99984 & 1 & 1 & 1 \\ & & & & 0.99803 & 1 & 1 & 1
\\ & & & &
1.07508 & 43/40 & 1.075 & 1 \\ & & & & 1.09886 & 43/40 & 1.075 & 2 \\
\hline
\end{tabular}
\end{center}
\caption{Scaling dimensions and multiplicities for the dilute $A_4$ model
in the $u<0$ branches.  The data for branch 4 agree very well with the
expansion \refeq{prodexp} of $(A_3,A_4)\times(A_2,A_3)$, and the data for
branch
3 agree reasonably with the expansion of $(A_4,A_5)\times(A_2,A_3)$.}
\label{sdA434}
\end{table}

\begin{table}[htbp]
\begin{center}
\begin{tabular}{|l|c|l|c||l|c|l|c|} \hline
\multicolumn{4}{|c||}{Branch 1} & \multicolumn{4}{c|}{Branch 2} \\ \hline
Approx. &
\multicolumn{2}{c|}{Exact} & Mult. & Approx. &
 \multicolumn{2}{c|}{Exact} & Mult. \\
 \hline 0.095234 & 2/21 & 0.095238\ldots & 2 & 0.13333 & 2/15 & 0.133\ldots
& 2 \\
0.285711 & 2/7 & 0.285714\ldots & 1 & 0.798273 & 4/5 & 0.8 & 1 \\ 0.951915
& 20/21 &
0.952381\ldots & 2 & 1.1335 & 17/15 & 1.133\ldots & 4 \\ 1.09586 & 23/21 &
1.09524\ldots
& 4 & 1.32242 & 4/3 & 1.33\ldots & 2 \\ 1.28583 & 9/7 & 1.28571\ldots & 2 &
1.79266 & 9/5
& 1.8 & 2 \\
\hline
\end{tabular}
\end{center}
\caption{Scaling dimensions and multiplicities for the dilute $D_4$ model
in the $u>0$ branches.  Comparison with the partition function expansions
\refeq{normexp} shows branch 1 to correspond to $(D_4,A_6)$, and branch 2 to
$(A_4,D_4)$; these are respectively tricritical and critical 3-state Potts.}
\label{sdD412}
\end{table}
\begin{table}[htbp]
\begin{center}
\begin{tabular}{|l|c|l|c||l|c|l|c|} \hline
\multicolumn{4}{|c||}{Branch 3} & \multicolumn{4}{c|}{Branch 4} \\ \hline
Approx. &
\multicolumn{2}{c|}{Exact} & Mult. & Approx. &
\multicolumn{2}{c|}{Exact} & Mult. \\
 \hline 0.09603 & 2/21 & 0.09524\ldots & 2 & 0.12450 & 1/8 & 0.125 & 1 \\
0.12520 & 1/8 &
0.125 & 1 & 0.13327 & 2/15 & 0.13333\ldots & 2 \\ 0.24414 & 37/168 &
0.22024\ldots & 1 &
0.25841 & 31/120 & 0.25833\ldots & 2 \\ 0.24414 & 37/168 & 0.22024\ldots &
1 & 0.80006 &
4/5 & 0.8 & 1 \\ 0.28376 & 2/7 & 0.28571\ldots & 1 & 0.92528 & 37/40 &
0.925 & 1 \\
 & & & & 0.99971 & 1 & 1 & 1 \\
 & & & & 1.11698 & 9/8 & 1.125 & 2 \\
 & & & & 1.13514 & 17/15 & 1.13333\ldots & 2 \\
\hline
\end{tabular}
\end{center}
\caption{Scaling dimensions and multiplicities for the dilute $D_4$ model
in the $u<0$ branches.  Branch 4 agrees well with the expansion
\refeq{prodexp} of
$(A_4,D_4)\times(A_2,A_3)$, and branch 3 agrees reasonably with that of
$(D_4,A_6)\times(A_2,A_3)$.}
\label{sdD434}
\end{table}

The estimates of the scaling dimensions $x_n$ allow the first few terms of the
isotropic
modular invariant partition functions to be determined, since these partition
functions are simply
\be
Z(q)=q^{-\frac{c}{12}}\round{1+\sum_{n=1}^{\infty}d_nq^{x_n}},
\ee
where $d_n$ are the multiplicities.
For the dilute models, we see
that branches 1 and 2 are described by the series of partition functions in
Table~\ref{MIPFs} and that the partition functions in branches 3 and 4 are
the product of
the critical Ising partition function
$(A_2,A_3)$ with those of branches 1 and 2 respectively.

Dilute $A_3$ therefore has the following modular invariant partition
functions in its
four branches:
$$
\begin{array}{ll}
\mbox{Branch 1:} & (A_3,A_4) \\
\mbox{Branch 2:} & (A_2,A_3) \\
\mbox{Branch 3:} & (A_3,A_4)\times (A_2,A_3) \\
\mbox{Branch 4:} & (A_2,A_3)\times (A_2,A_3)
\end{array}
$$ Similarly, dilute $A_4$ has the partition functions
$$
\begin{array}{ll}
\mbox{Branch 1:} & (A_4,A_5) \\
\mbox{Branch 2:} & (A_3,A_4) \\
\mbox{Branch 3:} & (A_4,A_5)\times (A_2,A_3) \\
\mbox{Branch 4:} & (A_3,A_4)\times (A_2,A_3)
\end{array}
$$ and dilute $D_4$,
$$
\begin{array}{ll}
\mbox{Branch 1:} & (D_4,A_6) \\
\mbox{Branch 2:} & (A_4,D_4) \\
\mbox{Branch 3:} & (D_4,A_6)\times (A_2,A_3) \\
\mbox{Branch 4:} & (A_4,D_4)\times (A_2,A_3)
\end{array}
$$
The isotropic expansions of these modular invariant partition functions
are, to the
relevant order,
\begin{eqnarray}
(A_2,A_3): && Z(q)=q^{-\frac{1}{24}}\square{1+\q{1}{8}+q+2\,\q{9}{8}+
                       4\,q^2+\Order{\q{17}{8}}} \nonumber \\
(A_3,A_4): && Z(q)=q^{-\frac{7}{120}}\square{1+\q{3}{40}+\q{1}{5}+
                       \q{7}{8}+2\,\q{43}{40}+3\,\q{6}{5}+\Order{\q{15}{8}}}
\nonumber \\
(A_4,A_5):
&& Z(q)=q^{-\frac{1}{15}}\square{1+\q{1}{20}+\q{2}{15}+
                         \q{1}{4}+\q{4}{5}+3\,\q{21}{20}+\Order{\q{17}{15}}}
\label{normexp} \\
(A_4,D_4): && Z(q)=q^{-\frac{1}{15}}\square{1+2\,\q{2}{15}+\q{4}{5}+
                          4\,\q{17}{15}+2\,\q{4}{3}+4\,\q{9}{5}+\Order{q^2}}
\nonumber \\
(D_4,A_6): && Z(q)=q^{-\frac{1}{14}}\square{1+2\,\q{2}{21}+\q{2}{7}+
                  2\,\q{20}{21}+4\,\q{23}{21}+2\,\q{9}{7}+\Order{\q{10}{7}}}
\nonumber
\end{eqnarray}
The products of each of these with the Ising partition function
$Z_I=(A_2,A_3)$ yield
\samepage{
\begin{eqnarray}
Z_I\times(A_2,A_3): \!\!&&\!\!
Z(q)=q^{-\frac{1}{12}}\square{1+2\,\q{1}{8}+\q{1}{4}+
                 2\,q+6\,\q{9}{8}+4\,\q{5}{4}+\Order{q^2}} \nonumber \\
Z_I\times(A_3,A_4): \!\!&&\!\!
Z(q)=q^{-\frac{1}{10}}\square{1+\q{3}{40}+\q{1}{8}+
                 2\,\q{1}{5}+\q{13}{40}+\q{7}{8}+q+\Order{\q{43}{40}}}
\nonumber \\
Z_I\times(A_4,A_5): \!\!&&\!\!
Z(q)=q^{-\frac{13}{120}}\square{1+\q{1}{20}+\q{1}{8}+
                \q{2}{15}+\q{7}{40}+\q{1}{4}+\q{31}{120}+\Order{\q{3}{8}}}
\label{prodexp} \\
Z_I\times(A_4,D_4):
\!\!&&\!\! Z(q)=q^{-\frac{13}{120}}\square{1+\q{1}{8}+2\,\q{2}{15}+
                2\,\q{31}{120}+\q{4}{5}+\q{37}{40}+q+2\,\q{9}{8}+
                \Order{\q{17}{15}}} \nonumber \\
Z_I\times(D_4,A_6): \!\!&&\!\!
Z(q)=q^{-\frac{19}{168}}\square{1+2\,\q{2}{21}+\q{1}{8}+
                2\,\q{37}{168}+\q{2}{7}+\Order{\q{23}{56}}} \nonumber
\end{eqnarray}
}
We find that the results for the two-colour models are similar to those of
the dilute
models. The modular invariant partition functions are all found to
be a product of two partition functions in Table~\ref{MIPFs} with the same
central
charge.  Thus, for example, the partition functions of the two-colour model
built on the
graph $A_4\otimes A_4$ are given by
$$
\begin{array}{ll}
\mbox{Branch 1:} & (A_4,A_5)\times (A_4,A_5) \\
\mbox{Branch 2:} & (A_3,A_4)\times (A_3,A_4)
\end{array}
$$
Our numerical estimates for the scaling dimensions of this model are
summarized in
Table~\ref{sdA4A412}.
The expansions of the isotropic partition function products are
\begin{eqnarray*}
\mbox{Branch 1:} \!\!&&\!\!
Z(q)=q^{-\frac{2}{15}}\square{1+2\,\q{1}{20}+\q{1}{10}+
                       2\,\q{2}{15}+2\,\q{11}{60}+2\,\q{1}{4}+\q{4}{15}+
                       2\,\q{3}{10}+\Order{\q{23}{60}}} \\
\mbox{Branch 2:} \!\!&&\!\!
Z(q)=q^{-\frac{7}{60}}\square{1+2\,\q{3}{40}+\q{3}{20}+
                       2\,\q{1}{5}+2\,\q{11}{40}+\q{2}{5}+2\,\q{7}{8}+
                       2\,\q{19}{20}+\Order{\q{43}{40}}}
\end{eqnarray*}

\begin{table}[htbp]
\begin{center}
\begin{tabular}{|l|c|l|c||l|c|l|c|} \hline
\multicolumn{4}{|c||}{Branch 1} & \multicolumn{4}{c|}{Branch 2} \\ \hline
Approx. &
\multicolumn{2}{c|}{Exact} & Mult. & Approx. &
\multicolumn{2}{c|}{Exact} & Mult. \\
 \hline 0.0499 & 1/20 & 0.05 & 2 & 0.0750 & 3/40 & 0.075 & 2 \\ 0.1016 &
1/10 & 0.1 & 1 &
0.1500 & 3/20 & 0.15 & 1 \\ 0.1333 & 2/15 & 0.133\ldots & 2 & 0.1996 & 1/5
& 0.2 & 2 \\
0.1877 & 11/60 & 0.1833\ldots & 2 & 0.2750 & 11/40 & 0.275 & 2 \\ 0.2489 &
1/4 & 0.25 & 2
& 0.4000 & 2/5 & 0.4 & 1 \\ 0.2798 & 4/15 & 0.266\ldots & 1 & 0.8839 & 7/8
& 0.875 & 2 \\
0.3107 & 3/10 & 0.3 & 2 & 0.9412 & 19/20 & 0.95 & 2 \\
 & & & & 1.0807 & 43/40 & 1.075 & 4 \\
\hline
\end{tabular}
\end{center}
\caption{Scaling dimensions and multiplicities for the two-colour
$A_4\otimes A_4$ model.  These agree well with the partition function
expansions of $(A_4,A_5)\times(A_4,A_5)$ in branch 1 and
$(A_3,A_4)\times(A_3,A_4)$ in branch 2.}
\label{sdA4A412}
\end{table}

Similarly, in the case of the model built on $D_4\otimes D_4$, the modular
invariant
partition functions are given by
$$
\begin{array}{ll}
\mbox{Branch 1:} & (D_4,A_6)\times (D_4,A_6) \\
\mbox{Branch 2:} & (A_4,D_4)\times (A_4,D_4)
\end{array}
$$ and in the case of the model built on $A_5\otimes D_4$ by
$$
\begin{array}{ll}
\mbox{Branch 1:} & (A_5,A_6)\times (D_4,A_6) \\
\mbox{Branch 2:} & (A_4,A_5)\times (A_4,D_4)
\end{array}
$$
All of our numerical results are consistent with the conjectured modular
invariant
partition functions summarized in Section~1.

\pagebreak

\section{Conclusion}

We have presented numerical evidence that the critical dilute \ade lattice
models in the
$u>0$ branches provide a realization of both the $(A,G)$ and the
$(G,A)$ series of modular invariant partition functions in the classification
of
Cappelli, Itzykson and Zuber.  In the
$u<0$ branches we have seen that the modular invariant partition functions are
products of the
$(A_2,A_3)$ partition function with members of the $(A,G)$ and $(G,A)$ series.
Furthermore, we have seen that the modular invariant partition functions of the
two-colour \ade lattice models at criticality are squares of members of the
$(A,G)$ and $(G,A)$ series.

Since these are all exactly solvable models, it would be interesting to see
some exact calculations of scaling dimensions to compare with the predicted
values.  Indeed such calculations have been performed in the case of the dilute
$A$ models in references \cite{PearZ95,PearZ295}.

We have here considered only unitary minimal conformal field theories.  It
would
also be interesting to see whether, by varying the crossing parameter, the
dilute \ade lattice models might provide realizations of non-unitary minimal
conformal field theories.


\section*{Acknowledgements}
\noindent We thank Jean-Bernard Zuber for pointing out Roche's conjecture
\refeq{conjecture}, Omar Foda for numerous discussions, and Ole Warnaar for
helpful suggestions.  This research is supported by the Australian Research
Council.

\goodbreak

\end{document}